\documentclass[prl,twocolumn,amsmath,amssymb]{revtex4}
\pdfoutput=1
\usepackage{graphicx}
\usepackage{color}
\usepackage{graphicx}% Include figure files
\usepackage{dcolumn}% Align table columns on decimal point
\usepackage{bm}% bold math
\usepackage{amsmath}
\usepackage{amsfonts}
\usepackage{bbm}
\usepackage{subfigure}
\usepackage{setspace}
\usepackage{pxfonts}
\usepackage{pstricks}

\newcommand{\beq}{\begin{eqnarray}}
\newcommand{\eeq}{\end{eqnarray}}

\newcommand{\bmp}{\noindent\begin{minipage}{16cm}}
\newcommand{\emp}{\end{minipage}\vskip 7mm} % 7mm untightened

\def\drawbox#1#2{\hrule height#2pt
        \hbox{\vrule width#2pt height#1pt \kern#1pt
              \vrule width#2pt}
              \hrule height#2pt}

\def\Asym#1#2{\vcenter{\vbox{\drawbox{#1}{#2}
              \kern-#2pt % line up boxes
              \drawbox{#1}{#2}}}}

\begin{document}
%%%%%%%%%%%%%%%%%%%%%%%%%%%%%%%%%%%%%%%%%%%%%%%%%%%%%%%%%%%%%%%%%%%%%%%%%%%
\title{\Large  \color{red}  Extra Electroweak Phase Transitions from Strong Dynamics}
\author{ $^{\color{blue}{\varheartsuit}}$Matti {\sc J\"arvinen}}
\email{mjarvine@ifk.sdu.dk}
\author{$^{\color{blue}{\clubsuit}}$Thomas A. {\sc Ryttov}}
\email{ryttov@nbi.dk}
\author{ $^{\color{blue}{\varheartsuit}}$Francesco {\sc Sannino}}
\email{sannino@ifk.sdu.dk}
 \affiliation{ $^{\color{blue}{\varheartsuit}}$Center for High Energy Physics, University of Southern Denmark, Campusvej 55, DK-5230 Odense M, Denmark. \\
$^{\color{blue}{\clubsuit}}$Niels Bohr Institute, Blegdamsvej 17, DK-2100 Copenhagen, Denmark }

%%%%%%%%%%%%%%%%%%%%%%%%%%%%%%%%%%%%%%%%%%%%%%%%%%%%%%%%%%%%%%%%%%%%%%%%%%%%%%%%%%%%%%%%%%%%%%%%%%%%%%%%%%%%%%%%%%%%%%%%%%%%%%%%%%%%%%%%%%%%%%

%%%%%%%%%%%%%%%%%%%%%%%%%%%%%%%%%%%%%%%%%%%%%%%%%%%%%%%%%%%%%%%%%%%%%%%%%%%%%%%%%%%%%%%%%%%%%%%%%%%%%%%%%%%%%%%%%%%%%%%%%%%%%%%%%%%%%%%%%%%%%%

\begin{abstract}
We show that models of dynamical electroweak symmetry breaking can possess an extremely rich finite temperature phase diagram. 
We suggest that early-universe extra electroweak phase transitions can appear in these models. 
\end{abstract}

%%%%%%%%%%%%%%%%%%%%%%%%%%%%%%%%%%%%%%%%%%%%%%%%%%%%%%%%%%%%%%%%%%%%%%%%
\maketitle
Models of dynamical electroweak symmetry breaking (DEWB) of the type summarized in \cite{Sannino:2008ha} are gaining momentum. Interesting applications have been envisioned both for the LHC phenomenology \cite{Sannino:2004qp,Foadi:2007ue} as well as Cosmology \cite{Nussinov:1985xr,Foadi:2008qv,Ryttov:2008xe,Nardi:2008ix}. Despite some initial studies  \cite{Cline:2008hr,Kikukawa:2007zk,Jarvinen:2009pk} the EW phase transition (EWPT) is still an uncharted territory. Understanding the DEWB at finite temperature may be relevant to explain the experimentally observed baryon asymmetry which could be 
generated at the EWPT
\cite{Shaposhnikov:1986jp,Gavela:1993ts,Nelson:1991ab}.  {An essential condition for EW
baryogenesis  to work is that its phase transition (PT) is strongly first order.
%(see for example
%ref.~\cite{Cline:2006ts} and references therein). 
In the Standard Model (SM) this condition is not satisfied 
 \cite{Kajantie:1995kf}. This provides an incentive for seeing whether the situation
improves in various extensions of the SM. 

Here we consider models of DEWB possessing a surprisingly rich finite temperature phase diagram structure. The basic ingredients are: i) Two different composite Higgs sectors; ii) One charged under the EW symmetry; iii) An underlying strong dynamics mixing the two sectors. An explicit realization just appeared in the literature \cite{Ryttov:2008xe} where we used new analytic results of \cite{Ryttov:2007cx}. These types of models  were envisioned earlier by  Eichten  and Lane \cite{Lane:1989ej}. 

We consider an asymptotically free gauge theory having sufficient matter to posses, at least, two independent non-abelian global symmetries spontaneously breaking, in the infrared, to two subgroups. One of the initial symmetries (or both) must contain the EW one in order to drive EW symmetry breaking. The Goldstones which are not eaten by the longitudinal components of the weak gauge bosons receive masses from other, unspecified, sectors.  Our analysis is sufficiently general that we need not to specify such sectors.

We denote with $\rm I$ and ${\rm I\,I}$ the two non-abelian global symmetries. They are broken at low temperatures and restored at very high temperatures. The restoration of each symmetry will typically happen at two different critical temperatures. We indicate with $\langle H_{\rm I}\rangle$ and  $\langle H_{\rm I\,I}\rangle$ the thermal average of the two condensates. The zero temperature physical masses $M_{{\rm I}}$ and $M_{{\rm I\,I}}$ of the two composite Higgses together with $\beta$ (measuring the mixing between the two), as well as the collection of all the other couplings mixing the two sectors constitute the parameters allowing us to make a qualitative picture of the complex phase structure. In the end we will confront our expectations with an explicit computation in a given model.

In figure ~\ref{one} we present three possible versions of the two-dimensional phase diagram as function of the temperature as well as one of the zero-temperature masses of one of the Higgses (holding fixed the other). The three plots are meant for three different strengths $\beta$ of the mixing while keeping the other relevant parameters fixed.  Four distinct regions are classified via the broken versus unbroken number of global symmetries. To simplify the discussion we are taking $\beta$ to be the parameter controlling the mixing between the two sectors. In fact, one should use the entire ensemble of parameters whose associated operators mix the different sectors. 

Let us describe the situation before embedding the EW symmetry within any of the two non-abelian global symmetries. We envision the following possibilities: i) The two sectors do not talk to each other ($\beta=0$). In this case the two PTs happen at different temperatures and do not interfere (left panel). ii) The two sectors do feel each other when $\beta\neq 0$. Possible phase diagrams are depicted in the central and right panel of Fig.~\ref{one}. 
\begin{figure}[htp!]
\includegraphics[height=2.5cm,width=2.6cm]{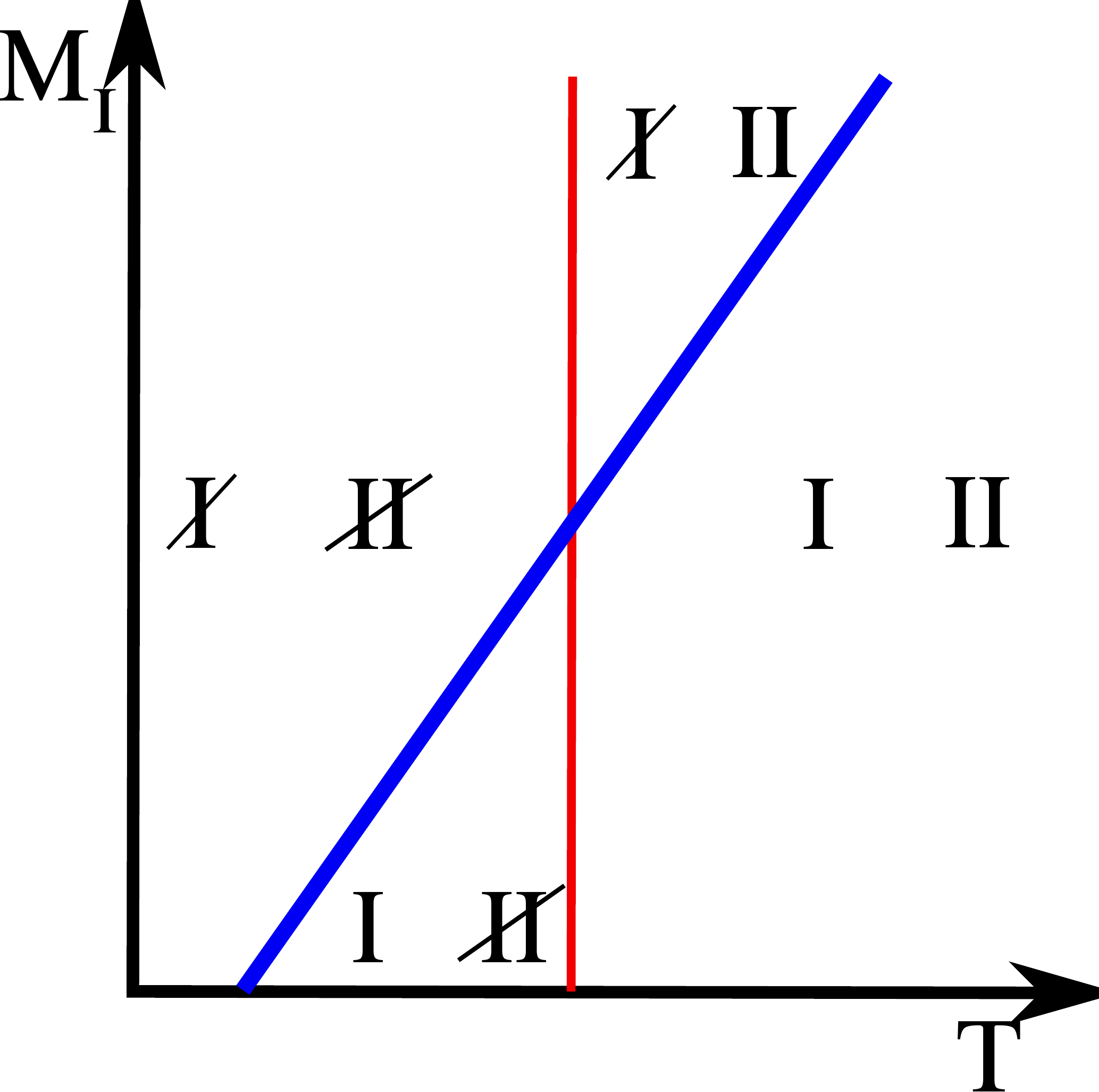} \hskip .2cm\includegraphics[height=2.5cm,width=2.6cm]{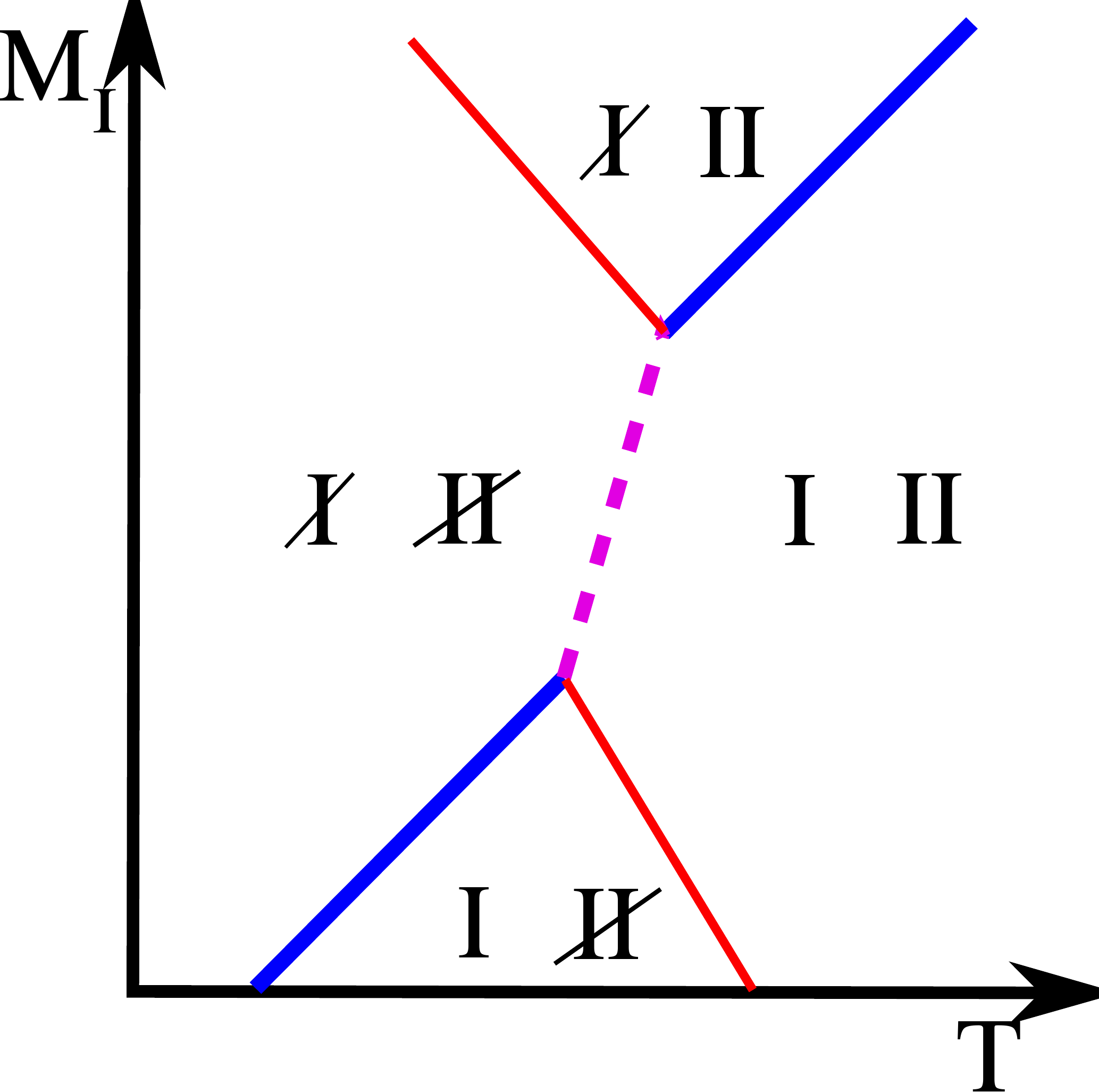}\hskip .2cm\includegraphics[height=2.5cm,width=2.6cm]{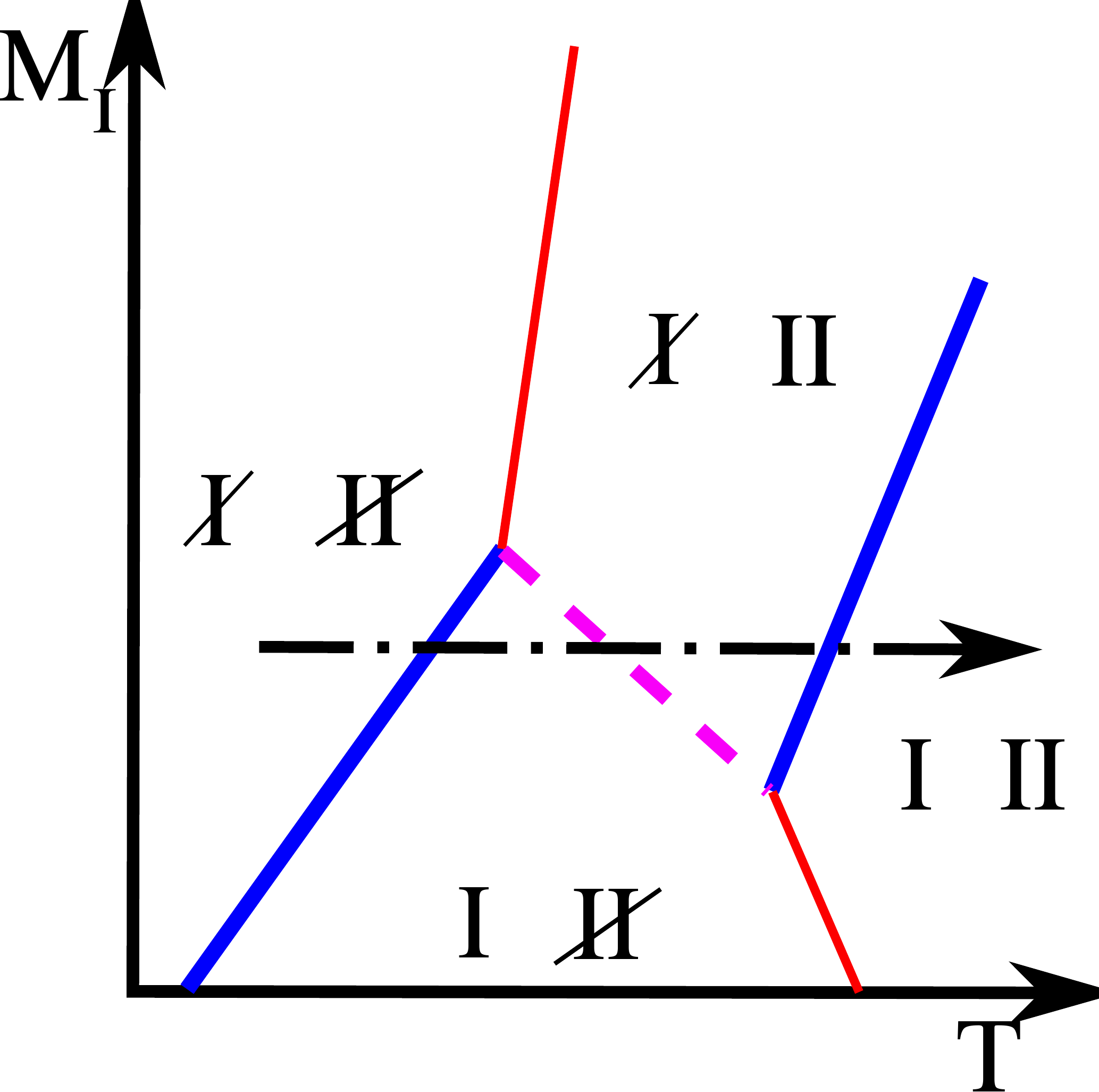}
\caption{Possible Phase Diagrams: {\bf Left Panel:} The two transitions do not feel each other ($\beta=0$). {\bf Central and Right Panels:} The two transitions do interfere with each other  ($\beta \neq 0$).  }\label{one}
\end{figure}
In a generic strongly coupled theory the two global symmetries are bound to talk to each other and hence the second possibility is the one expected. A new line can develop (the dashed one depicted in the central and right panel) entirely due to the interactions between the two sectors. This line allows several new possible PTs. {}For example, according to the central phase diagram the transition between two broken to two unbroken phases can occur at the same critical temperature along the dashed line in the $M_{\rm I}-T$ plane. What strikes us as a very intriguing possibility is the pattern of PTs one can encounter following the right panel phase diagram. Along the horizontal dashed arrow line we have three subsequent PTs constituted by the first condensate being melted twice and re-generated once while the second one melts only once. We could also plot a diagram similar to the one in the right panel but with the first vertex lower than the second one (with respect to $M_{\rm I}$).  In fact, more sophisticated PTs can occur.

We turn on the EW fields by gauging the relevant symmetries within, for definitiveness, the first sector. We do not have a SM Higgs but require the new strong dynamics to drive EW symmetry breaking.  The units of the dynamically generated scale of the new strong dynamics are now fixed by the mass of the weak gauge bosons. Would the {\it extra} transition associated to the right-panel diagram survive? What are the main effects of the SM on the phase diagram?  It would be very interesting if a complex PT structure appears in technicolor-like extensions of the SM when the universe reaches temperatures near the EW scale. Similar possibilities have been investigated earlier in the case of the two Higgs doublet model \cite{Land:1992sm}. The EW fields will impinge on the PTs and the relevant degrees of freedom are the weak gauge bosons and the SM fermions. The gauge fields couple via covariant derivatives while the fermions communicate by means of effective Yukawa-type interactions  as proposed in Minimal Walking Technicolor (MWT) \cite{Sannino:2004qp,Foadi:2007ue}. In fact, the top quark has the major impact on the phase diagram due to its very large Yukawa coupling.

To provide quantitative answers to the questions raised above we have used as a template the Ultra Minimal Walking Technicolor (UMT) model \cite{Ryttov:2008xe}. 
We stress that UMT is used here only as an explicit example. Therefore, we do not attempt to discuss in detail the finite temperature analysis here.
We only summarize the results supporting the expectations of the possibility of {\it extra} EWPTs within technicolor models with several technimatter representations \cite{Lane:1989ej,Sannino:2004qp}. We provide the full technical details in \cite{Jarvinen:2009pk} where we study quantitatively the strengths of the transitions and cover a wide region of the parameter space of the theory.

We use the zero temperature linear effective Lagrangian describing the relevant low energy degrees of freedom associated to the underlying UMT gauge dynamics consisting of an $SU(2)$ technicolor gauge theory with two types of underlying matter fermions: Two Dirac flavors in the fundamental representation of the gauge group and one Dirac flavor in the Adjoint representation. The two relevant non-abelian global classical symmetries are then $SU(4)$ and $SU(2)$ which are both expected to break spontaneously, in the vacuum and at zero temperature, to $Sp(4)$ and $SO(2)$ respectively. In addition there is an anomaly free abelian $U(1)$ global symmetry under which all the fermions are charged. The effective Lagrangian is 
\begin{eqnarray}
\mathcal{L} &=& \frac{1}{2} \text{Tr}\left[ D_{\mu} N_{\rm I} D^{\mu}N_{\rm I}^{\dagger} \right] + \frac{1}{2} \text{Tr}\left[ \partial_{\mu} N_{\rm I\,I} \partial^{\mu}N_{\rm I\,I}^{\dagger} \right] \nonumber \\
&& - \mathcal{V}\left( N_{\rm I}, N_{\rm I\,I} \right) + \mathcal{L}_{ETC} \ ,
\end{eqnarray}
where the scalar fields are
$N_{\rm I}  = \left[ \frac{1}{2}\left( H_{\rm I} + i\Theta_{\rm I} \right) + \sqrt{2} \left( i\Pi_{\rm I}^i + \tilde{\Pi}_{\rm I}^i \right) X_{\rm I}^i \right]E_{\rm I}$ and $N_{\rm I\,I} = \left[ \frac{1}{\sqrt{2}}\left( H_{\rm I\,I} + i\Theta_{\rm I\,I} \right) + \sqrt{2} \left( i\Pi_{\rm I\,I}^i + \tilde{\Pi}_{\rm I\,I}^i \right) X_{\rm I\,I}^i \right]E_{\rm I\,I}$ with 
\begin{eqnarray}
E_{\rm I}  = \left( \begin{array}{cc}
0_{2\times 2} & 1_{2\times 2} \\
-1_{2\times 2} & 0_{2\times 2}
      \end{array} \right) \ , \qquad 
E_{\rm I\,I}  = \left( \begin{array}{cc}
0 & 1 \\
1 & 0
      \end{array} \right) \ .
\end{eqnarray}
The low-energy spectrum consists of the two composite Higgs particles $H_{\rm I}$ and $H_{\rm I\,I}$ together with their associated pseudoscalar partners $\Theta_{\rm I}$ and $\Theta_{\rm I\,I}$. The Goldstone bosons appearing due to the breaking of the global symmetries are denoted by $\Pi_{\rm I}^i,\ i=1,\ldots,5$ and $\Pi_{\rm I\,I}^i,\ i=1,2$ while $\tilde{\Pi}_{\rm I}^i,\ i=1,\ldots,5$ and $\tilde{\Pi}_{\rm I\,I}^i,\ i=1,2$ are their associated scalar partners. Also $X_{\rm I}^i,\ i=1,\ldots,5$ and $X_{\rm I\,I}^i,\ i=1,2$ are the broken generators for which an explicit realization can be found in \cite{Ryttov:2008xe}.

As discussed above the EW gauge group is embedded in $SU(4)$ only. It gives rise to the following covariant derivative \cite{Ryttov:2008xe}
\begin{eqnarray}
D_{\mu}N_{\rm I}&=& \partial_{\mu}N_{\rm I} -i\left[ G_{\mu} N_{\rm I} + N_{\rm I}  G_{\mu}^{T} \right] \ , \label{covariant1}\\
G_{\mu} &=& \left( \begin{array}{cc}
gW_{\mu}^a \frac{\tau^a}{2} & 0 \\
0 & -g' B_{\mu} \frac{\tau^3}{2}
\label{covariant}
\end{array} \right) \ ,
\end{eqnarray}
where $g$ and $g'$ are the EW gauge couplings while $W_{\mu}^a,\ a=1,\ldots,3$ and $B_{\mu}$ are the EW gauge bosons. The potential of the theory is chosen to preserve the anomaly free $SU(4) \times SU(2) \times U(1)$ global symmetry and reads: 
\begin{eqnarray}
%\mathcal{V} &=&
 &&\Big\{-\frac{m_{\rm I}^2}{2} \text{Tr} \left[ N_{\rm I} N_{\rm I}^{\dagger} \right] + \frac{\lambda_{\rm I}}{4} \text{Tr} \left[ N_{\rm I} N_{\rm I}^{\dagger} \right]^2 +\lambda_{\rm I}' \text{Tr} \left[ N_{\rm I} N_{\rm I}^{\dagger} N_{\rm I} N_{\rm I}^{\dagger} \right] 
%\right. 
\nonumber \\
&& %\left. 
+ \left( {\rm I}\rightarrow {\rm I\,I} \right) \Big\}  + \frac{\delta}{2} \text{Tr}\left[ N_{\rm I}N_{\rm I}^{\dagger} \right] \text{Tr} \left[ N_{\rm I\,I} N_{\rm I\,I}^{\dagger} \right]
 \nonumber \\
&&  %\nonumber \\
%&& + 
+4 \delta' \left[ \left( \det N_{\rm I\,I} \right)^2 \text{Pf} N_{\rm I} + \text{h.c.} \right] \ . 
\end{eqnarray} 
Here $\text{Pf}N_{\rm I}$ is the Pfaffian. The $\delta$ and $\delta'$ terms allow for the $SU(4)$ and $SU(2)$ sectors to communicate with each other. In the limit $\delta' \rightarrow 0$ the symmetry is enhanced to $U(4) \times U(2)$.  $\beta$ is given by
\begin{eqnarray}
\tan \left( 2\beta \right) &=& \frac{2v_{\rm I\,I}v_{\rm I}\left( 2\delta'v_{\rm I\,I}^2  -\delta \right)}{m_{\rm I\,I}^2 - m_{\rm I}^2 - \delta v_{\rm I}^2  - \left(  \delta' v_{\rm I\,I}^2 - \delta  \right)v_{\rm I\,I}^2} \ ,
\end{eqnarray}
where $v_{\rm I}$ and $v_{\rm I\, I}$ are the zero temperature VEV's of $H_{\rm I}$ and $H_{\rm I\,I}$ found by minimizing the above potential. One should also note that in the limit $\delta,\delta' \rightarrow 0$ we have $\beta =0$. We finally remark on $\mathcal{L}_{ETC}$ which denotes a set of terms giving mass to the Goldstone bosons not eaten by the longitudinal components of the weak gauge bosons as well as some of the other composite states. For an explicit realization see \cite{Ryttov:2008xe}.

The specific values of $M_{\rm I}$, $M_{\rm I\,I}$ and $\beta$ depend on the underlying gauge dynamics. What we investigate here is, in effect, the phase diagram of the effective Lagrangian per se while the intrinsic UMT dynamics will have to be unveiled via first principles lattice simulations. We do not assume the various transitions to be second order, but if they are then we could use the Wilson approach as done, for example, in \cite{Sannino:2004ix}. However, we are interested (for baryogenesis purposes) in understanding the strength of the PTs which cannot be estimated within the Wilson approach. Hence, we use the effective potential method to study the phase diagram.
We employ one-loop high temperature approximation together with the
summation of the higher order ring-diagrams and with the finite temperature masses for the EW gauge bosons %to evaluate the effective potential
in our numerical calculations following the pioneering work in \cite{Carrington:1991hz}. We identify a significant region
of parameter space where the ratio of the composite Higgs vacuum expectation value to the critical temperature $\phi_c/T_c \gtrsim 1$ for either of the transitions as required by electroweak baryogenesis \cite{Jarvinen:2009pk}. 
%this ratio is sufficiently large to induce electroweak baryogenesis. 
We have checked the validity of the high-$T$ expansion for the other regions of our plots by adding higher order terms in the expansion and seeing how the results change. Including terms up to
and including order $1/T^6$, we find that the quantitative results
presented here are stable against higher order corrections. A similar analysis has been performed in the simpler case of MWT \cite{Cline:2008hr}. 

\vskip .3cm
UMT has an axial anomaly which destroys one of the two $U(1)$s and its effects are encoded in the $\delta'$ term. We have studied the two cases $\delta' = 0$ and $\delta' \neq 0$. In this letter we concentrate on the former case. We have checked that the $\delta'$ term does change the details of the phase diagram, however, it still allows for a similar rich structure. 
%Taking $\delta'=0$ corresponds to having an extra Goldstone %in the theory. 
\begin{figure}[htp!]
{\includegraphics[height=2.9cm,width=2.9cm]{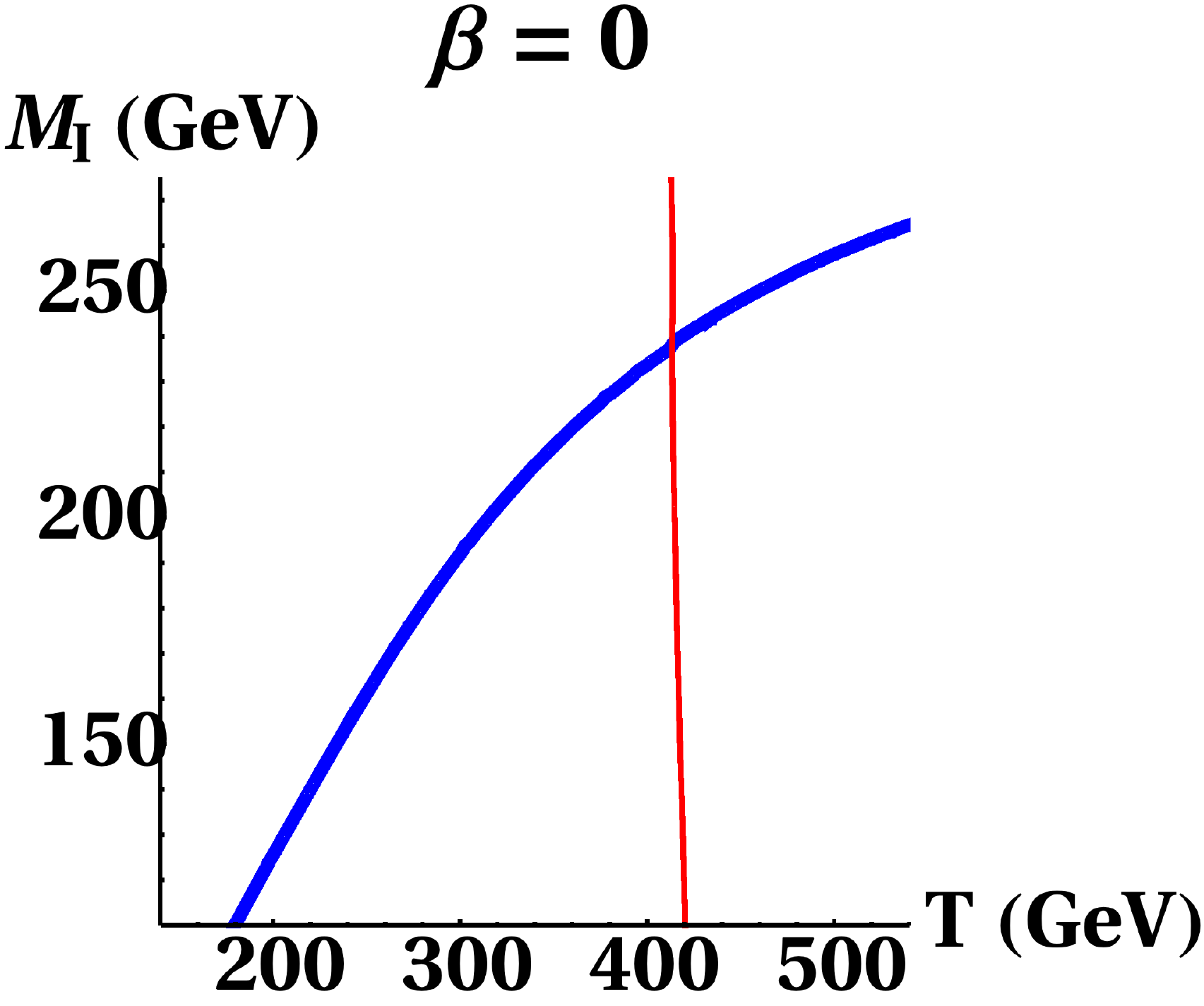}%\hskip .1cm 
\includegraphics[height=2.9cm,width=2.9cm]{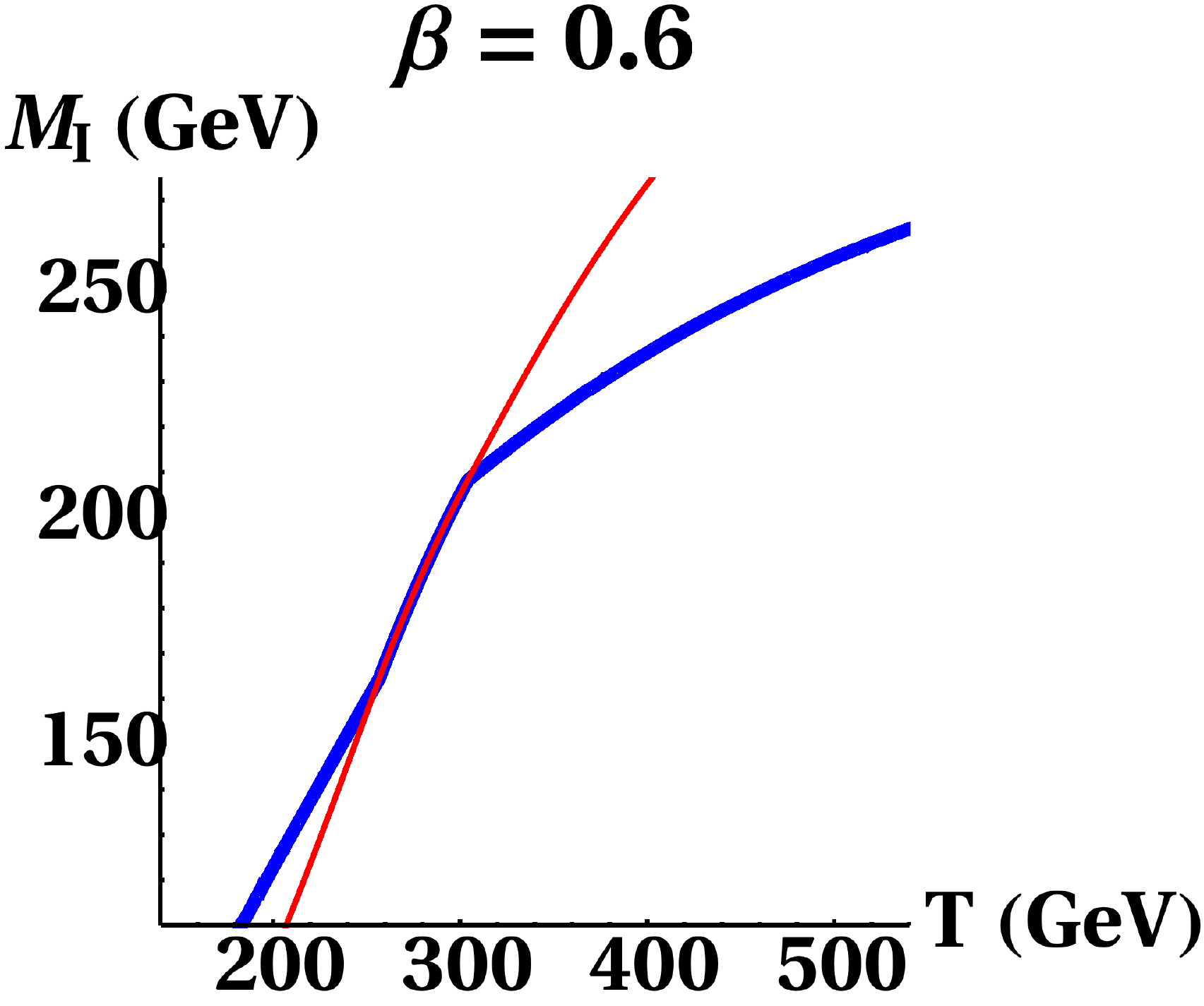}%\hskip .1cm
\includegraphics[height=2.9cm,width=2.9cm]{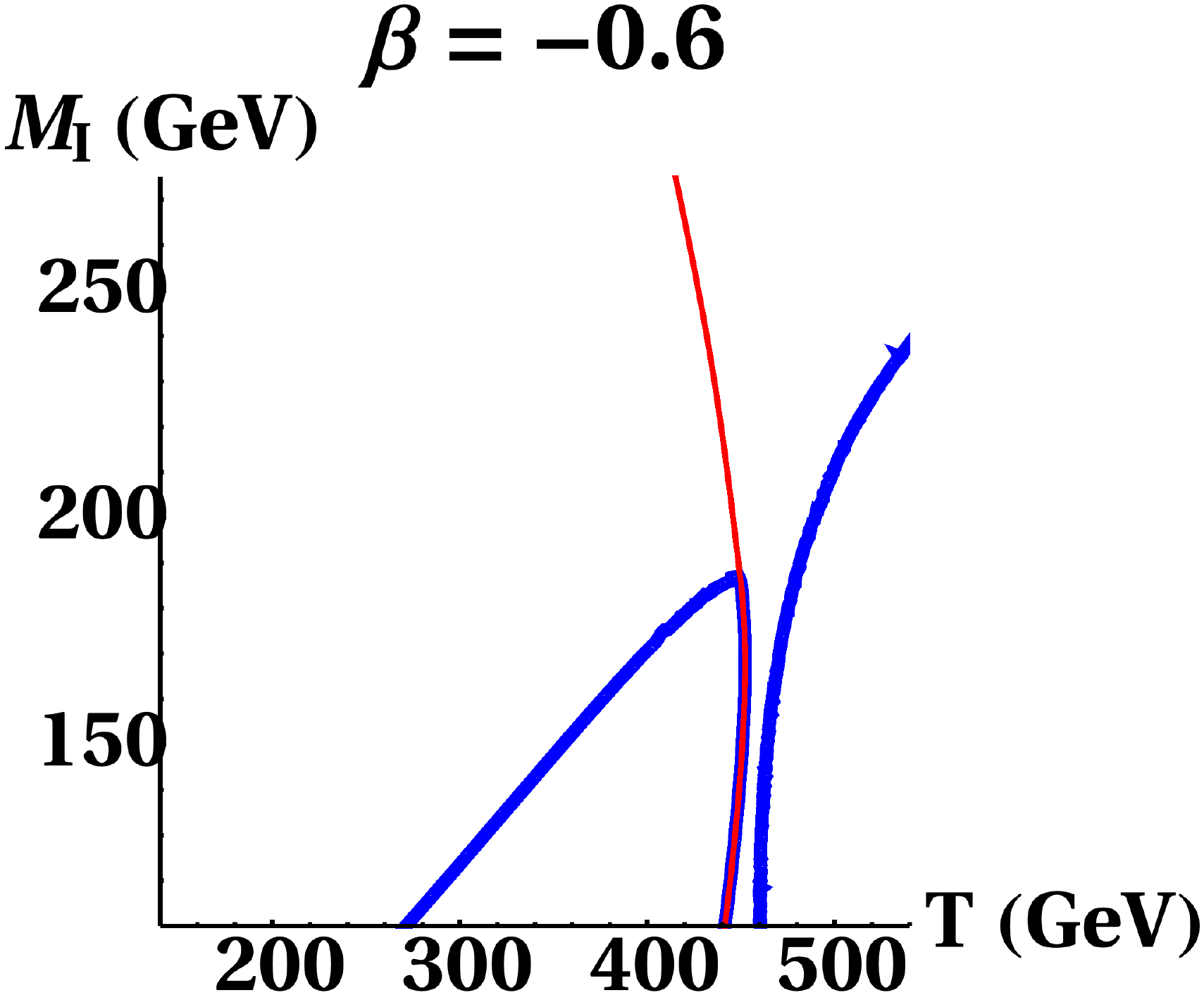}}
\caption{Possible Phase Diagrams in UMT without EW: {\bf Left Panel:} The two transitions do not feel each other ($\beta=0$). {\bf Central  ($\beta = 0.6 $) and Right ($\beta = -0.6 $) Panels:} The two transitions do interfere with each other.}\label{fig:b}
\end{figure}

Let us first study the phase diagram before embedding the EW interactions and without the Yukawa terms.
We indicate the $SU(4)\rightarrow Sp(4)$, (sector ${\rm I}$) transition with a thick (blue) line and the $SU(2)\rightarrow SO(2)$ transition (sector ${\rm I\, I}$) by a thin (red) line. We present three different phase diagrams in Fig.~\ref{fig:b} for three different values of $\beta$. 
The left panel in Fig.~\ref{fig:b} corresponds to the case of no mixing among the two sectors ($\beta= 0$ i.e. $\delta=0$). The remaining  parameters are encoded in the zero-temperature values of the physical masses of the different degrees of freedom such as the $\tilde{\Pi}_{{\rm I}}$ and $\tilde{\Pi}_{{\rm I\,I}}$ and ${\Theta}$ and $\tilde{\Theta}$.  The specific range of the parameters we use to plot the phase diagram is such that: i) At least one of the two sectors features a strong PT (the ${\rm I\, I}$ sector); ii) The two critical temperatures are near to each other; iii) The global phase diagram (for $\beta \neq 0$)  shows a strong interplay between the two sectors. We  take
$M_{\tilde{\Pi}_{\rm I}} \simeq 150$~GeV and  $M_{\tilde{\Pi}_{\rm I\,I}} \simeq 500$~GeV. $M_{\tilde{\Theta}}$  and $M_{\Theta}$ are both zero because of the two unbroken $U(1)$s.  

We find the thin red transition to be strongly first order while the thick blue one is first order for very small $M_{\rm I}$ and  ends into a second order point around $M_{\rm I} \simeq  130$ GeV when the $M_{\tilde{\Pi}_{\rm I} }$ is taken to be around $150$~GeV. In the plot we kept fixed $M_{\rm I\,I}$ at around $280$~GeV and $\langle H_{\rm I\,I}\rangle \simeq 300$ GeV. The energy units are obtained imposing that the zero temperature VEV of $H_{\rm I}$ (once the theory is EW gauged) drives the EW breaking and hence its zero temperature value is $246$~GeV. 
%We will, however, consider the effects of the EW gauging and the %coupling to the SM fermions below.  

{}For $\beta \neq 0$ the two sectors communicate as it can be deduced from the central and right panel of Fig.~{\ref{fig:b}}. In this case the two transitions meet on a first order line in the $M_{\rm I}-T$ plane. The right panel shows the {\it extra} transition occurring in the range $100 <M_{\rm I} <190$~GeV. The order and strength of the $\rm I$ PT away from the region in which the two PTs coalesce is affected by the chosen value of the remaining parameters of the low energy effective theory and here it is second order. On the coalescing line it is first order.

Due to the interplay (natural in strongly coupled gauge theories) between the two different sectors we find the following general results relative to the phase diagram: 
 {i) A region, in the phase diagram, of simultaneous (same critical temperature) first order PT occurs;}
{ii) A region on the phase diagram appears where $\langle H_{\rm I} \rangle$ first melts and then {\it regenerates} at the critical temperature point associated to the melting of the second condensate and finally  it melts again at an even higher temperature leading to the intriguing phenomenon of {\it extra} PTs.} 
The presence of the extra PT occurs for a negative value of $\beta$. 

\vskip .3cm
{ How does the embedding of the EW and Yukawa sector of the SM affect the phase diagram discussed above?} 
 
We introduce the EW gauge bosons by gauging the $SU(2)\times U(1)$ subgroup of $SU(4)$ \cite{Ryttov:2008xe} (see Eq.~\eqref{covariant1}) and also endow the SM fermions with a mass term by introducing effective Yukawa operators featuring the composite $SU(4)$ Higgs \cite{Foadi:2007ue}. The presence of the new physical states substantially alters the finite temperature effective potential. The reader can find a detailed account of the effects of these terms in a similar computation specialized to the case of MWT  \cite{Cline:2008hr}. The most dramatic effect is due to the top Yukawa interaction. 

We find that when using the same parameters chosen for plotting the phase diagram, in the absence of the EW, the phase diagram region featuring the {\it extra} PT shrinks. 
The plots in Fig. \ref{fig:c} show a phase diagram (only qualitatively) similar to the one presented above. We see that the two transitions can still substantially affect each other. This, however, occurs for $M_{\rm I\, I}\simeq 150$~GeV rather than $280$~GeV (keeping fixed $\langle H_{\rm I \,I} \rangle \simeq 300$~GeV), with an overall physical mass (including the ETC dynamics) of $\tilde{\Pi}_{\rm I}$ around $380$~GeV,  $\tilde{\Pi}_{\rm I\,I}$ mass $520$~GeV and $\delta'=0$. 

With the parameters chosen, the first order line common to both transitions develops at a higher value of the Higgs ($H_{\rm I}$) mass. 
\begin{figure}[htp!]
{\includegraphics[height=2.9cm,width=2.9cm]{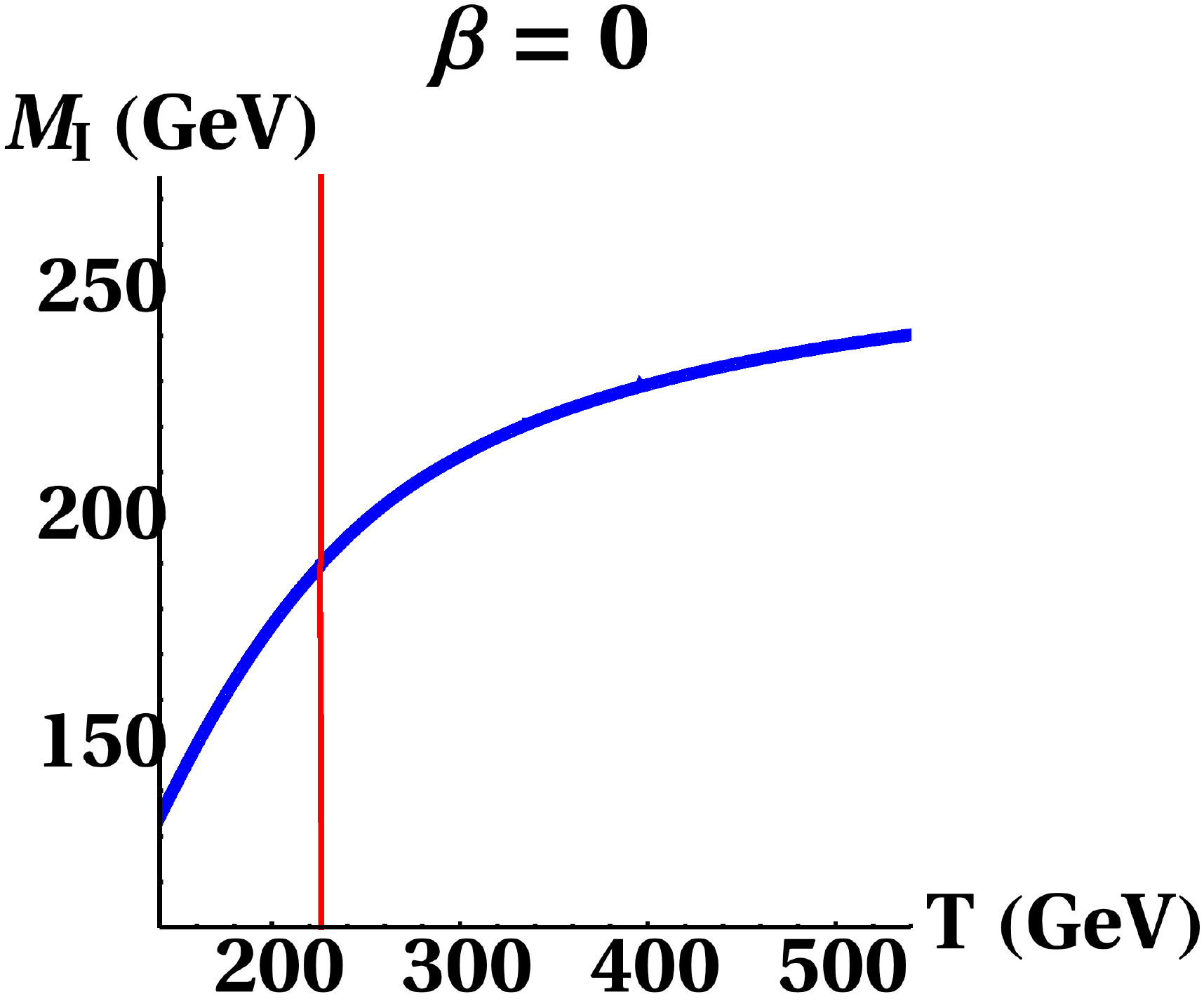}%\hskip .1cm
 \includegraphics[height=2.9cm,width=2.9cm]{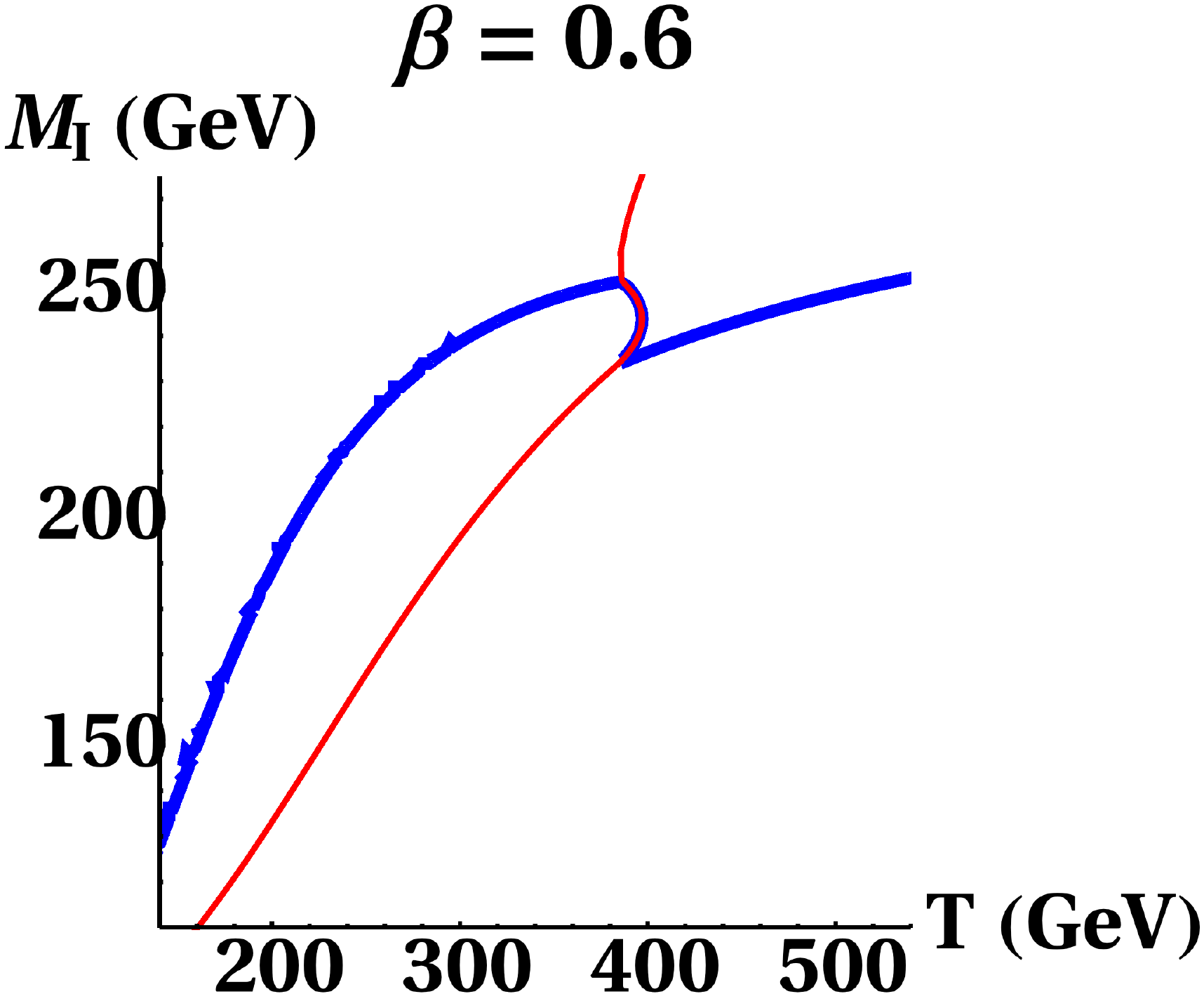}%\hskip .1cm
\includegraphics[height=2.9cm,width=2.9cm]{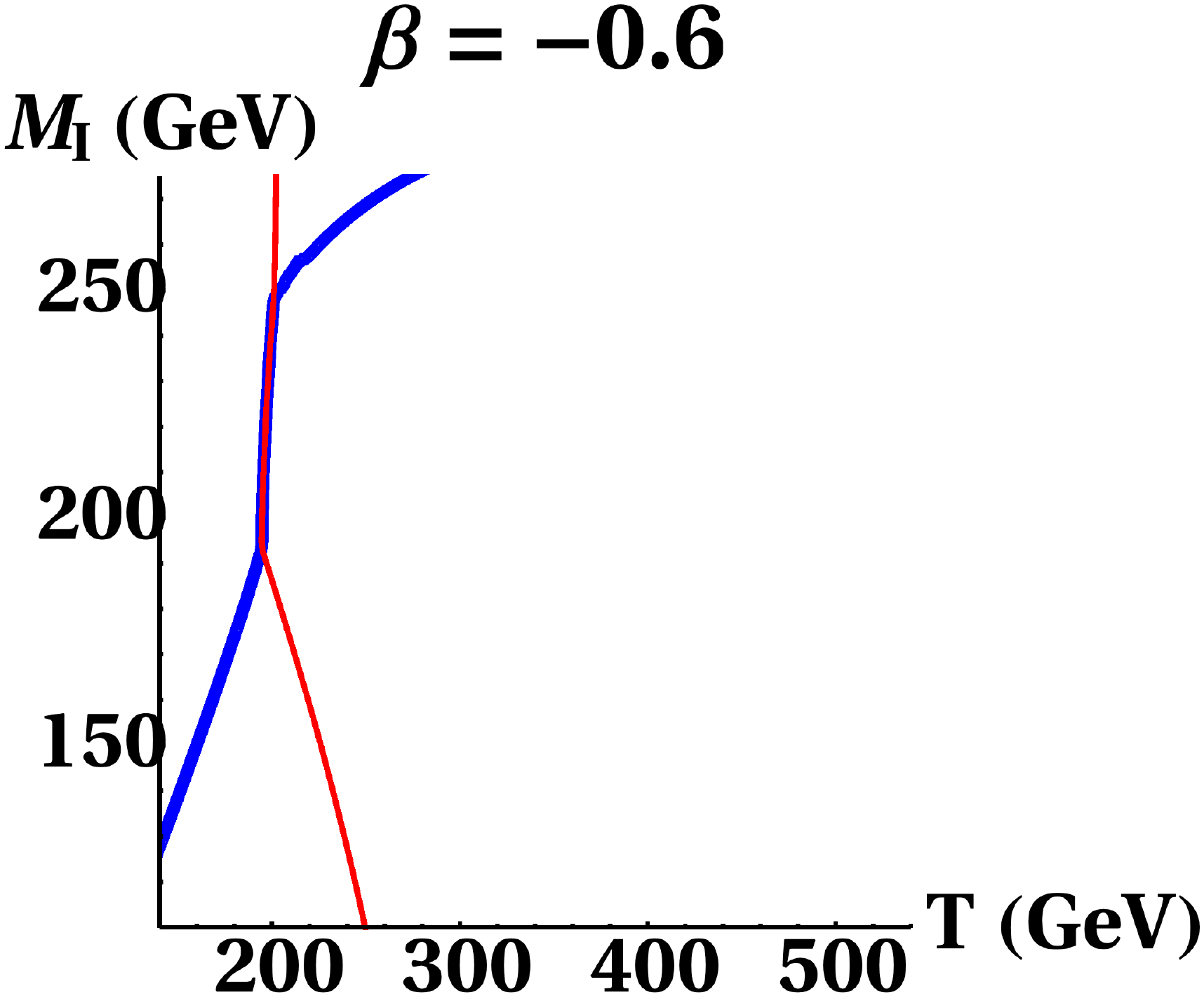}}
\caption{Possible Phase Diagrams in UMT with EW: {\bf Left Panel:} The two transitions do not feel each other ($\beta=0$). {\bf Central  ($\beta = 0.6 $) and Right ($\beta = -0.6 $) Panels:} The two transitions do interfere with each other. }\label{fig:c}
\end{figure}
We also observe that the EWPT (thick-blue line) occurs at a lower value of the critical temperature. This phenomenon is due to the combined effect of the top and $\tilde{\Pi}$ corrections. 
The left panel on the left of Fig. \ref{fig:c}  shows the little or no interplay between the two transitions, the center panel plot shows the interesting case of an {\it extra} EWPT as well as the possibility of further delay the PT. This now occurs for a positive $\beta$ because of the very light $M_{\rm I \,I}$. The right panel plot shows no evidence of an {\it extra} PT. The coalescence lines for the plots in the central and right panels correspond to simultaneous first order PTs for the two sectors. 

We showed that models of DEWB possess an extremely rich finite temperature phase diagram. We demonstrated that {\it extra} EWPTs can appear in a general class of models of which UMT is an explicit example.  It would be interesting to investigate the associated gravitational spectrum (see for example \cite{Delaunay:2007wb}). The interplay of the EWPT with the center group symmetry \cite{Mocsy:2003qw,Sannino:2008ha,Cline:2008hr} of the underlying technicolor theory -- intimately related to the confinement physics of the new dynamics -- will lead to an even richer phase diagram.

\end{document}